\documentclass[11pt,a4paper]{article}
\usepackage{jinstpub}

\usepackage{graphicx}% Include figure files
\usepackage{dcolumn}% Align table columns on decimal point
\usepackage{bm}% bold math
\usepackage{wasysym}
\usepackage{subfig}
\usepackage{multirow}
\usepackage{mathtools}
\usepackage[mathlines]{lineno}% Enable numbering of text and display math
\title{Using Machine Learning to Select High-Quality Measurements}
\author[a,1]{Andrew Edmonds,\note{Now at Boston Unversity, Boston, USA}}
\author[a]{David Brown,}
\author[b]{Luciano Vinas}
\author[c]{and Samantha Pagan}

\affiliation[a]{Lawrence Berkeley National Laboratory, Berkeley, California 94720, USA}
\affiliation[b]{University of California, Berkeley, California 94720, USA}
\affiliation[c]{University of North Carolina, Chapel Hill, North Carolina 27514, USA}
\emailAdd{awje@bu.edu}
\date{\today}% It is always \today, today,
\abstract{We describe the use of machine learning algorithms to select high-quality measurements for the Mu2e experiment. This technique is important for experiments with backgrounds that arise due to measurement errors. The algorithms use multiple pieces of ancillary information that are sensitive to measurement quality to separate high-quality and low-quality measurements.} %Training the algorithms on weighted events directs the algorithm to prioritize the identification of higher-weighted events and we find that weighting the lowest-quality measurements significantly improves the separation power of these algorithms.
\notoc
\begin{document}
\maketitle

\flushbottom

%\tableofcontents

\section{Introduction}
\label{sec:introduction}
Artificial neural networks (ANNs) and boosted decision trees (BDTs) are important tools in particle physics experiments. They have been used to identify single-top production at the Tevatron~\cite{Abazov:2008kt, Aaltonen:2010jr}, discover the Higgs at the LHC~\cite{Aad:2012tfa, Chatrchyan:2012ufa}, and distinguish neutrino flavors in oscillation experiments~\cite{Roe:2004na, AguilarArevalo:2007it}. In these examples, the machine learning algorithms separated signal processes from background processes by combining measurements of multiple variables that each had some small separation power. In this paper, we present an example of a different use for machine learning algorithms: to separate high-quality and low-quality measurements.

Separating high-quality and low-quality measurements is important for experiments whose signals and backgrounds arise from similar physics processes. In these experiments there may only be one physics variable that distinguishes signal from background. This means that a mismeasurement of this variable will lead to backgrounds being mistaken for signals and vice versa. By separating high-quality and low-quality measurements, we can select high-quality measurements to enhance the signal sensitivity of such experiments.%Using machine learning algorithms, we can exploit ancillary information sensitive to measurement quality to identify and remove low-quality measurements and prevent backgrounds being mistaken for signals.%To reduce such background events, low-quality measurements must be identified. 

Machine learning algorithms exploit ancillary information to separate different classes of events. For measurement quality, information such as the number of noise hits in the detector or the number of hits used in the fit will each offer some small separation power. By synthesizing the information from multiple variables, machine learning algorithms can achieve a larger separation. %Furthermore, by weighting measurements according to their quality during training, the machine learning algorithm prioritizes the identification of the weighted measurements and a greater separation power can be achieved.

Such techniques have been used before. The AMS experiment~\cite{Aguilar:2013qda} used a BDT to identify reconstructed tracks with an incorrectly assigned charge~\cite{Zuccon:2013tvl}. This paper demonstrates the same technique for the Mu2e experiment~\cite{Bartoszek:2014mya}.

%separate electrons and positrons and cut on the BDT output such that the cut ``minimizes the overall measurement of uncertainties''~\cite{Corti:2014ria}. In this paper, we take a more direct approach and train machine learning algorithms to separate measurements by their quality directly.

The outline of the rest of the paper is as follows. In section~\ref{sec:mu2e}, we briefly describe the Mu2e experiment and motivate why it needs machine learning algorithms to extract scientific results. In section~\ref{sec:input-variables}, we describe the measurement quality variables used in this study. In section~\ref{sec:training}, we describe the training procedure used to train the machine learning algorithms. In section~\ref{sec:results}, we present the results of using the trained algorithms to separate high-quality and low-quality measurements. We conclude in section~\ref{sec:summary}.

\section{The Mu2e Experiment}
\label{sec:mu2e}
\begin{figure}[!htbp]
  \centering
    \includegraphics[width=1.0\textwidth]{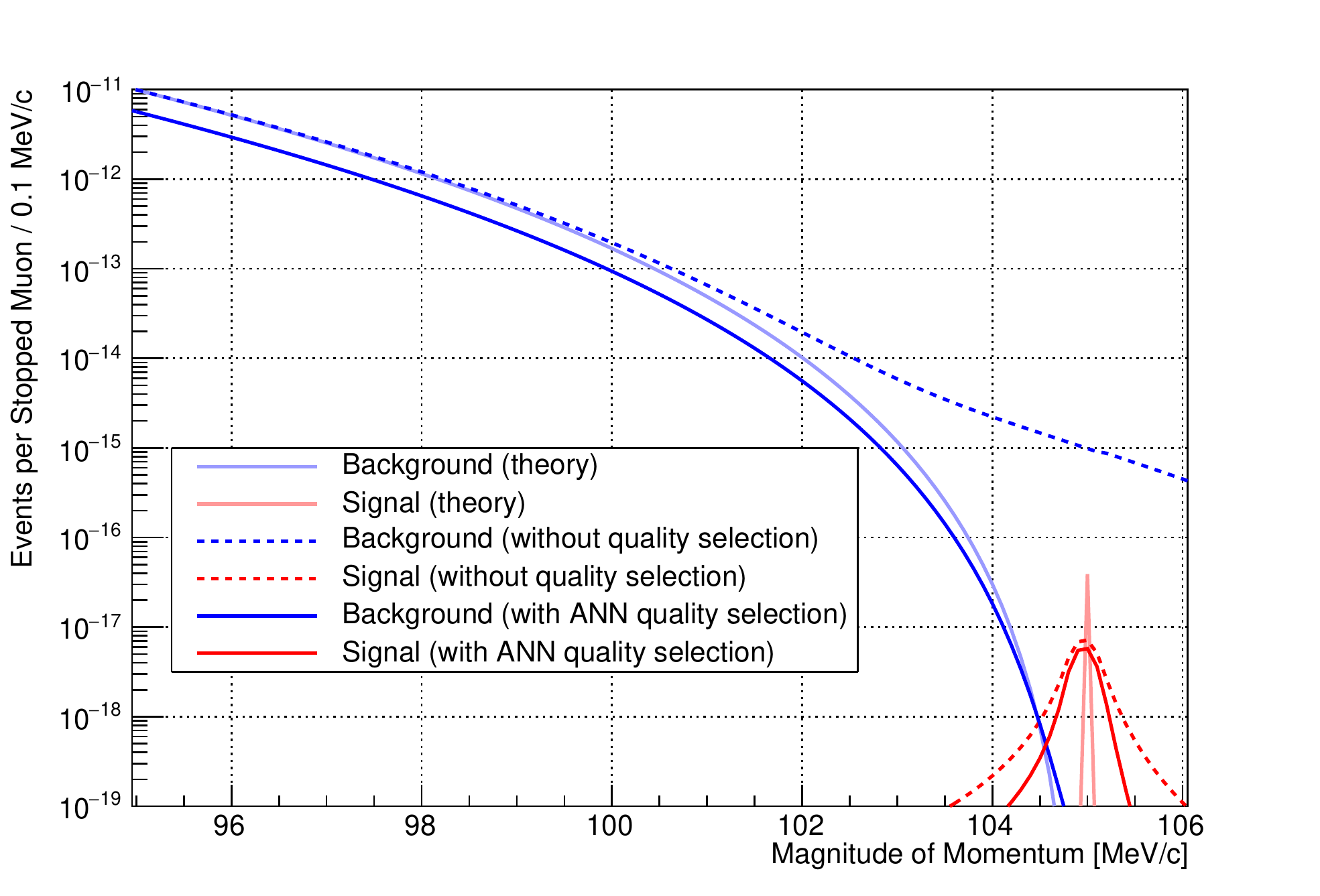}
  \caption{Energy spectra of signal electrons (red) and background electrons (blue) for various scenarios. The pale curves show the theoretical spectra~\cite{Czarnecki:2011mx, Szafron:2016cbv} $\left(\text{assuming signal rate is } 10^{-16} \right)$, the dashed curves show the theoretical spectra convolved with the measurement resolution without a measurement quality selection (figure~\ref{fig:results:mom-res}, black), and the solid curves show the theoretical spectra convolved with the measurement resolution with the ANN measurement quality selection (figure~\ref{fig:results:mom-res}, blue).}
  \label{fig:signal-dio}
\end{figure}

%\subsection{Overview}
%($\mu\rightarrow e$), ($\mu\rightarrow e \nu \bar{\nu}$).
% ($\mu^{-} + \text{Al} \rightarrow  e^{-} + \text{Al}$) 
The Mu2e experiment~\cite{Bartoszek:2014mya} will search for the charged lepton flavor violating process of $\mu-e$ conversion by stopping muons in an aluminum stopping target. A stopped muon will emit a signal electron when it converts to an electron only, and a background electron when it decays to an electron and two neutrinos. The only variable that distinguishes the signal process from the background process is the momentum of the emitted electron. The momentum spectrum of signal electrons is monoenergetic at 105 MeV/c~\cite{Czarnecki:2011mx}, and the momentum spectrum of background electrons~\cite{Czarnecki:2011mx, Szafron:2016cbv} falls steeply and ends at 105 MeV/c (figure~\ref{fig:signal-dio}, pale colors). The Mu2e experiment will use a straw tube tracker to measure the momentum of electrons by reconstructing their helical trajectories in a solenoidal magnetic field. The resolution of the momentum measurement will smear the theoretical spectra and cause signal and background to overlap. %Mu2e will measure the momentum of electrons emitted by stopped muons.

%Signal electrons will have a momentum of 105 MeV/c and background electrons will have a range of momenta. The background spectrum falls steeply and ends at the signal momentum ().% shows the theoretical momentum spectra for the signal (with radiative corrections~\cite{ceLL}) and the background. %The momentum spectrum of the background is steeply-falling and ends at the $\mu-e$ conversion momentum.

\begin{figure}[!t]
  \centering
  \includegraphics[width=1.0\textwidth]{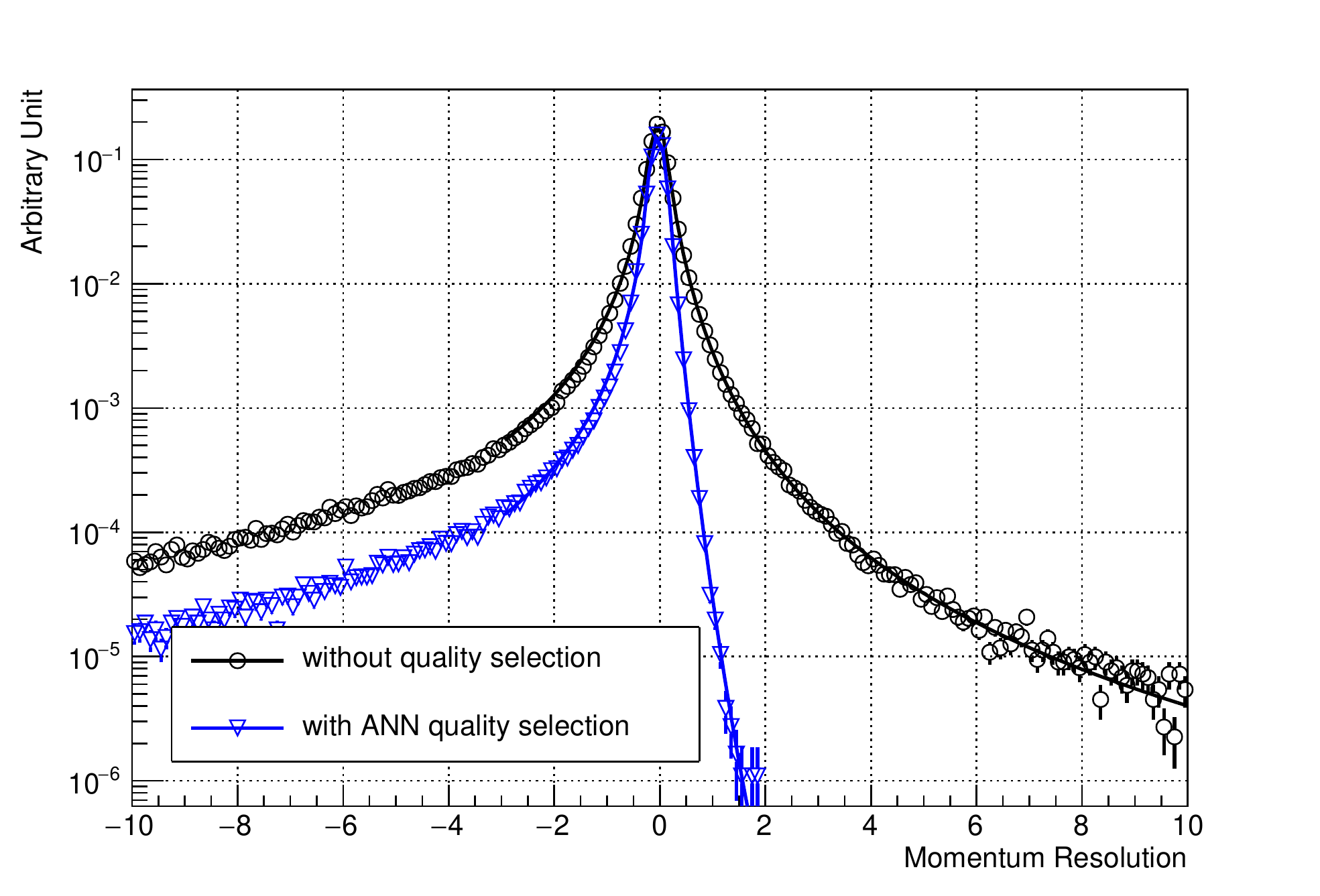}
  \caption{Reconstructed momentum resolution for simulated electrons with an ANN quality selection (blue triangles) and without a quality selection (black circles) and their associated double-sided Crystal Ball fits.}
  \label{fig:results:mom-res}
\end{figure}

% Without identifying low-quality measurements, Mu2e will not be able to observe a signal. 
To get the expected momentum resolution (figure~\ref{fig:results:mom-res}, black circles), we simulate and reconstruct high-energy (75 -- 110 MeV/c) electrons being emitted from the stopping target. The Mu2e simulation propagates the electrons through a realistic model of the experiment and calculates the energy deposited in individual detector elements with Geant4~\cite{Allison:2016lfl, Allison:2006ve, Agostinelli:2002hh}. Then, a dedicated tracker electronics simulation (tuned to prototype data) converts the energy deposited in the tracker straws into realistic tracker hits. To produce the expected hit environment for the track reconstruction, we add tracker hits from other processes (e.g.\ proton emission after nuclear muon capture or photon conversion in the stopping target). Standard Mu2e track finding identifies hits of potential signal tracks from the thousands of total hits per event, and a Kalman Filter fit reconstructs the electron trajectory from these hits, using simulated annealing~\cite{Kirkpatrick:1983zz} to improve the hit purity. From the Kalman Filter fit, we have a measurement of the electron's momentum. Comparing this to the Monte Carlo truth momentum of the electron gives us the momentum resolution.
% Simulated annealing is used to optimize the fit over non-Gaussian degrees of freedom, such as errors in pattern recognition and the sign ambiguity when translating drift times to position constraints in a drift.
The measurement resolution has a Gaussian core and non-Gaussian tails. The Gaussian core is due to multiple scattering of the electron, and the non-Gaussian tails are due to energy losses (on the low side) and the non-linear straw response, impact of pile-up hits, and pattern recognition errors (on the high side). The distribution fits well to a double-sided Crystal Ball function~\cite{Oreglia:1980cs} in the high-side tail, in the core and at the start of the low-side tail. The furthest part of the low-side tail is not important for our studies.

Figure~\ref{fig:signal-dio} (dashed) shows the convolution of the theoretical energy spectra with the double-sided Crystal Ball fit and we see that the Mu2e signal is overwhelmed by background. The large background comes from background electrons in the high-side tail of the resolution function. Although the fraction of such measurements is small, there are enough low-energy background electrons that this produces a significant challenge for the experiment. To overcome this challenge, Mu2e needs to identify and remove low-quality measurements in the high-side tail of the resolution. %By identifying and removing low-quality measurements, Mu2e can reduce the high-side tail of the momentum resolution and stop the signal from being smothered.

\section{Measurement Quality Variables}
\label{sec:input-variables}
The variables selected for this study each have some sensitivity to the measurement quality. From the Kalman fit, we know both the fraction and absolute number of hits that survived the simulated annealing process and are used in the fit $\left( f_{\text{used}},~n_{\text{used}} \right)$. Large values of these variables indicate a well-constrained fit with few background hits in the final reconstruction stage. Also from the Kalman fit, we can determine the tracker straws that \textit{should} have seen hits and calculate the fraction of straws that did contain hits $\left( f_{\text{expected}} \right)$. A large value would show that there are missing hits. The Kalman fit tries to assign a drift distance to the straw hits. If a large fraction of hits do not have an assigned drift distance $\left( f_{\text{drift}} \right)$, then that would indicate a larger uncertainty in the hit position. Finally, the Kalman fit reports the uncertainty in the momentum $\left( p_{\text{err}} \right)$ and time $\left( t_{\text{err}} \right)$ of the fit, and a fit chi-squared consistency $\left( con \right)$ can be calculated to determine how closely the fit matches the hits. Table~\ref{tab:input-variables} summarizes the input variables and orders them by their separations. Separation is defined in ref.~\cite{Hocker:2007ht} as:

\begin{equation}
  \label{eqn:separation}
  \left< S^{2} \right> = \dfrac{1}{2} \int \dfrac{\left( \hat{y}_{S}(y) - \hat{y}_{B}(y)\right)^{2}}{\hat{y}_{S}(y) + \hat{y}_{B}(y)} dy,
\end{equation}
where $\hat{y}_{S}$ and $\hat{y}_{B}$ are the signal and background PDFs of $y$. A separation of zero indicates identical PDFs, and a separation of one indicates PDFs with no overlaps.

\begin{table}[!htbp]
  \begin{center}
%    \resizebox{1.0\columnwidth}{!}{
    \begin{tabular}{|c|c|c|}%c|c|}
      \hline
      Input Variable & Brief Description & Separation\\% & ANN & BDT \\
      \hline
      $p_{\text{err}}$ & fit momentum error & 0.4919\\% & 1 & 2 \\
      $n_{\text{used}}$ & number of used hits & 0.4131\\% & 6 & 1 \\
      $f_{\text{expected}}$ & fraction of expected detector elements hit & 0.2122\\% & 2  & 4 \\
      $f_{\text{drift}}$ & fraction of hits without a drift distance & 0.1900\\% & 7  & 5 \\
      $f_{\text{used}}$ & fraction of used hits & 0.1305\\% & 5  & 3 \\
      $t_{\text{err}}$ & fit time error & 0.1274\\% & 3 & 7 \\
      $con$ & fit chi-square consistency & 0.09520\\% & 4 & 6 \\
      \hline
      \end{tabular}
%    } 
  \end{center}
  \caption{The input variables used for the machine learning algorithms ordered by separation~\cite{Hocker:2007ht}. The separation of a variable is independent of the specific training. A separation of zero is total overlap and a separation of one is total separation.}
  \label{tab:input-variables}
\end{table}

We selected the input variables for this study from an initial pool of potential quality-sensitive variables. From these variables, we only used those with a strong effect on the training. This was determined by comparing different ANNs with individual variables removed. Since each ANN had a different response value for a given track, comparing ANN response values between different ANNs was meaningless. We therefore renormalized the response values to follow a uniform distribution from 0 to 1 and selected the 77\% of tracks with the highest scores. To reach the value of 77\%, we optimized a reference ANN with all potential input variables to reject 99\% of low-quality measurements. With this technique we could check the resolution distribution of the highest-quality measurements and keep variables whose removal resulted in a larger high-side resolution tail (i.e. a worse resolution).% If the removal of the input variable resulted in a worse resolution, then that variable was considered a strong variable and kept in the training. %The input variables listed in Table~\ref{tab:input-variables} were all considered strong by this method.

\section{Training Procedure}
\label{sec:training}
We used TMVA~\cite{Hocker:2007ht} to train machine learning algorithms to separate high-quality and low-quality measurements. We used a large sample of simulated electrons with half used for training the algorithms and the other half used for testing the results.
%[maybe resolution figure here with signal and background? add it to input variable figure? add it to previous section?].

For the training, we defined high-quality measurements as those with a difference between the reconstructed measurement and the true value of less than 250 keV/c $\left( |p_{\text{reco}} - p_{\text{true}}| < 250~\text{keV/c} \right)$, and low-quality measurements as those in the high-side tail of the resolution where the behavior becomes non-Gaussian $\left( p_{\text{reco}} - p_{\text{true}}>700~\text{keV/c} \right)$.

%To improve the separation between high-quality and low-quality measurements, we apply an exponential weighting factor $w = \max \left(1.0,~e^{2.0\min\left(\text{reco - true},~3.0\right)} \right)$ to weight the measurements furthest in the high-side tail of the resolution as more important for the algorithm to separate.
%The weight increases exponentially for $\Delta p > 0$ MeV/c up to a maximum at $\Delta p = 3.0$ MeV/c. %By setting an upper limit, we avoid overtraining on very few events in the far tail.

%It was not possible to perform a box-cut optimization with a background weight expression and so we trained the ANN with and without this weighting.

We trained an ANN and a BDT. The ANN nodes had sigmoid activation functions, the network contained $\left( N,N-1 \right)$ nodes in the hidden layers, and it was trained for 500 cycles. The BDT used 850 trees with a maximum depth of 3, a minimum number of events per node of 150, and adaptive boosting~\cite{freund1997decision} with $\beta=0.5$. Preliminary studies varied these parameters and found the variations to have a negligible effect on the results. We also investigated weighting the background sample to train the algorithms to weight the worst measurements as most important to separate. This also had a negligible effect on the results.

To compare to a non-machine learning algorithm, we performed a box-cut optimization with TMVA. This optimization finds a set of cuts of the same input variables for 100 high-quality measurement efficiency bins.

The output distributions of the independent training and testing samples matched and so there is no evidence that these algorithms were overtrained.

\section{Results}
\label{sec:results}
The performance of a machine learning algorithm is summarized in its ``receiver operating characteristic'' (ROC) curve, which shows the low-quality measurement rejection power of the algorithm as a function of high-quality measurement efficiency.

\begin{figure}[!htbp]
  \centering
    \includegraphics[width=1.0\textwidth]{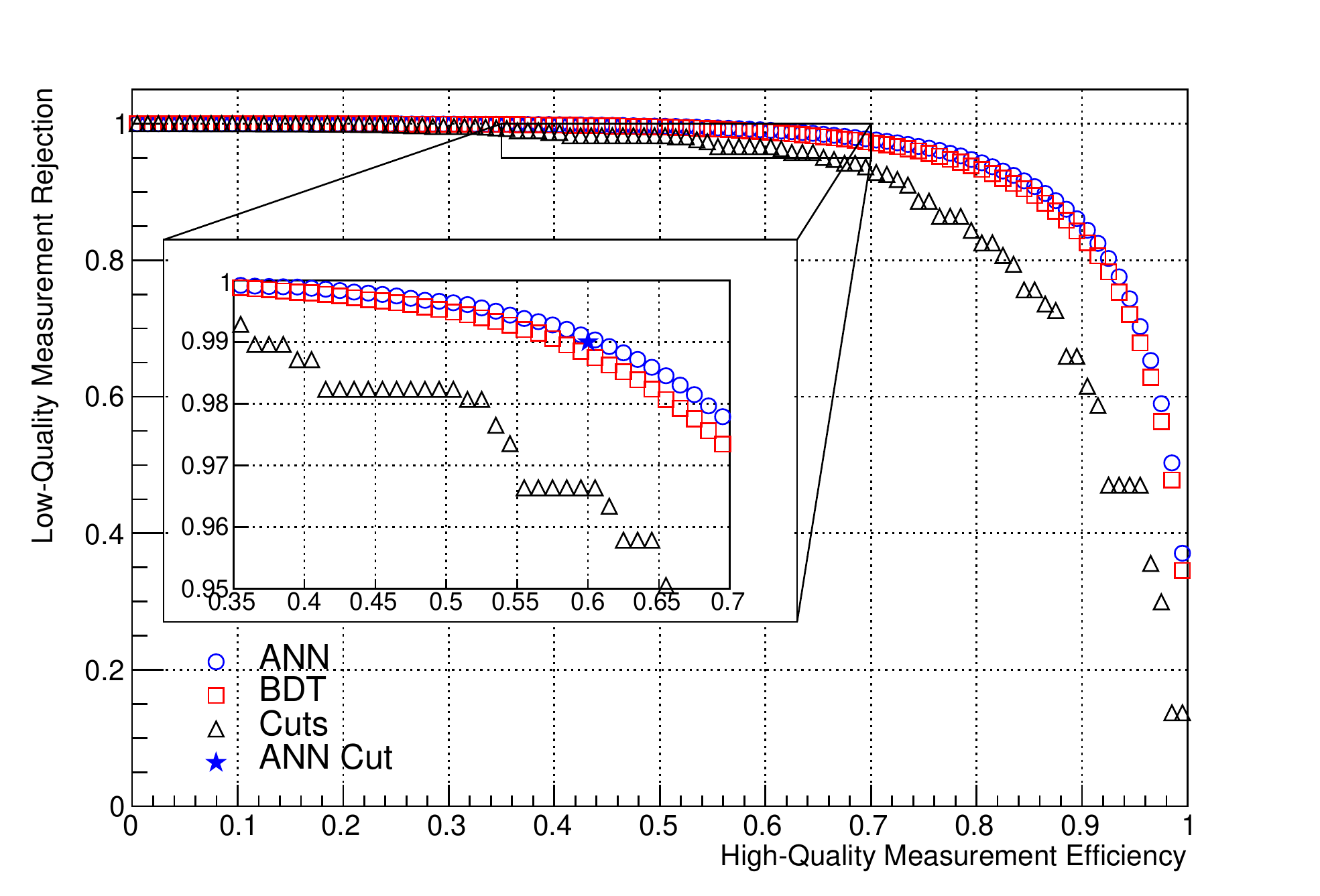}
  \caption{ROC curves for the BDT (red squares), ANN (blue circles) and box-cut optimization (black triangles). The ANN cut used for figure~\ref{fig:signal-dio} (solid) and figure~\ref{fig:results:mom-res} (blue) is shown as a blue star in the inset.}
  \label{fig:roc-curves}
\end{figure}

Figure~\ref{fig:roc-curves} shows the ROC curves of the ANN, BDT, and box-cut optimization. The ROC curves of the ANN and BDT (figure~\ref{fig:roc-curves}, blue circles and red squares, respectively) show similar performances (i.e.\ for a given low-quality measurement rejection, the high-quality measurement efficiency is similar), although we note that the BDT output value for each track takes four times longer to calculate than the ANN output value. The box-cut optimization (figure~\ref{fig:roc-curves}, black triangles) performs significantly worse than the machine learning algorithms (i.e.\ for a given low-quality measurement rejection, the high-quality measurement efficiency is lower).

We used each algorithm to select tracks such that we had a low-quality measurement rejection of 99\%. The high-quality measurement efficiency at this rejection power was 0.60 for the ANN, 0.57 for the BDT, and 0.35 for the box-cut optimization. Selecting measurements with this level of low-quality measurement rejection reduces the high-side tail of the momentum resolution (figure~\ref{fig:results:mom-res}, blue triangles), which leads to a much improved situation in the measured energy spectrum (figure~\ref{fig:signal-dio}, solid) where the signal peak is no longer overwhelmed by background. %With this selection, Mu2e can achieve its scientific goals~\cite{Bartoszek:2014mya}.

%To determine the physics impact for Mu2e, we selected high-quality measurements from the simulated Mu2e dataset using the background-weighted ANN so that we achieved a low-quality measurement rejection of 0.99 (figure~\ref{fig:roc-curves}, blue star). 

%% \begin{figure}[!htbp]
%%   \centering
%% %  \subfloat[][momentum resolution for tracks with a ``simple'' track quality selection (black) and the ANN track quality selection (red)\label{fig:results:mom-res}]{\includegraphics[width=1.0\columnwidth]{figs/MomRes_SimpleAndANN_wFit_forPaper.pdf}}\\
%%   %\subfloat[][expected momentum spectra for conversion electrons (red) and DIO electrons (blue) with ANN track quality selection (solid) and a ``simple'' track quality selection (dashed)\label{fig:results:mom}]{\includegraphics[width=1.0\columnwidth]{figs/DIOSpectrum_SmearedSimpleAndANN_wFit_forPaper.pdf}}
%%   %  \includegraphics[width=1.0\columnwidth]{figs/MomSpectra_CeDIO_TruthAndANN_forPaper.pdf}
%%   \includegraphics[width=1.0\columnwidth]{figs/MomSpectra_CeDIO_All_forPaper.pdf}
%%   \caption{Momentum spectra of signal electrons (red) and background electrons (blue) for both theory (dashed)\cite{Szafron:2016cbv, ceLL} and theory $\times$ resolution after measurement quality selection (solid).}
%%   \label{fig:results}
%% \end{figure}

\section{Summary}
\label{sec:summary}
In this paper, we described the training of machine learning algorithms to identify high-quality measurements. The improvement in high-quality measurement efficiency is a factor of 1.7 over a box-cut optimization for a low-quality measurement rejection of 99\%.

For Mu2e, the optimal choice of quality selection will be the result of an optimization process that takes into account all background processes to maximize signal sensitivity. The machine learning algorithms presented in this paper are an important tool for Mu2e because they minimize the high-side tail of the momentum resolution, which would produce a large background contribution if it were left untreated. Further, the technique described in this paper may be useful for other precision experiments.

\section{Acknowledgements}
The authors of this paper would like to thank our colleagues in the Mu2e collaboration for developing the Mu2e simulation and dedicated tracker electronics simulation that allowed us to perform this work. We would also like to acknowledge the support of the SULI program at LBNL.

\bibliographystyle{JHEP.bst}
\bibliography{main}

\end{document}